\documentclass[12pt]{iopart}
\usepackage{graphicx}
\expandafter\let\csname equation*\endcsname\relax
\expandafter\let\csname endequation*\endcsname\relax
\usepackage{amsmath}
\usepackage{amssymb}
\usepackage{cite}
\begin{document}

\title{Quantum super-resolution for imaging two pointlike entangled photon sources}
\author{Huan Zhang$^{1}$, Wei Ye$^{2}$, Ying Xia$^{1}$, Zeyang Liao$^{1*}$, Xue-hua Wang}
\address{1. State Key Laboratory of Optoelectronic Materials and Technologies, School of Physics, Sun Yat-sen University, Guangzhou 510275, China}
\address{2. School of Information Engineering, Nanchang Hangkong University, Nanchang 330063, China}
\ead{liaozy7@mail.sysu.edu.cn}

\vspace{10pt}

\begin{abstract}
We investigate the resolution for imaging two pointlike entangled sources by using the method of the moments and the spatial-mode demultiplexing (SPADE), where the pointlike entangled sources can be generated by injecting single-mode sources with arbitrary quantum statistics distribution into an optical parametric amplifier (OPA). We demonstrate that the separation estimation sensitivity is mainly determined by the photon distribution in each detected modes and it can be enhanced by either increasing the squeezed parameter of the OPA or eliminating the relative phase difference of the entangle sources.  Furthermore, in the limiting case of infinitely small source separation, the usage of  entangled sources can have better resolution than those using incoherent and coherent sources. The results here can find important applications for the quantum super-resolution imaging and quantum metrology.
\end{abstract}

\section{Introduction}
Due to the Abbe's diffraction limit \cite{1, 2}, the minimum resolvable separation of classical optical instrument is about half wavelength of the detection light source (i.e., $d_{min}=\lambda/2NA$, where $\lambda$ is the wavelength of light and $NA$ is the numerical aperture). How to further improve the resolution of the optical instrument is always a hot research topic which has attracted extensive interests in the past few decades  \cite{Vangindertael2018, Pujals2019}. In the past few decades, a number of methods have been proposed to overcome the diffraction limit, such as the stimulated emission depletion microscopy (STED) \cite{Hell2000,Chen2015}, structured illumination microscopy (SIM) \cite{Gustafsson2000,Mudry2012,Zeng2014,Xi2019}, the single-molecule localization microscopy (SMLM) \cite{Betzig2006,Zhuang2006}, and stochastic optical fluctuation imaging (SOFI) \cite{Dertinger2009}. These methods have been widely used for biological imaging with typical resolution being about 20-50 nm. Using quantum effects such as quantum entanglement \cite{Boto2000,Shih2001,Moreau2019,Cui2023}, quantum coherence \cite{Liao2010,Liao2012,Rui2016} and quantum statistics \cite{Cui2013,Classen2017, Tenne2019, Bhusal2022}, the diffraction limit can also be in principle overcome. A natural question arises that what is the ultimate limit of quantum imaging?

The above question may be addressed from the point of view of quantum metrology \cite{Delaubert2008,Tsang2015,Giovannetti2011, 26}. 
The simplest task for the superresolution optical imaging is discrimination of two close pointlike sources \cite{27,28,29,30,31,32,33,34}.  According to the theory of quantum metrology, the quantum limit for the separation estimation of point-like sources is determined by the quantum Cramer-Rao bound $(\Delta d)^2\leq1/F_Q$, where $F_Q$ is the quantum Fisher information (QFI), quantifying the sensitivity of the quantum state of the point source to the change of distance $d$ \cite{35,36,37,38}. Based on this idea, Tsang et al. showed that the estimation error is not diverging as the separation between the two point sources become infinitely small and moreover they proved that the spatial-mode demultiplexing method (SPADE) can saturate this quantum bound \cite{Tsang2015,27,29,30}. In particular, the separation of two equally bright light sources within the diffraction limit can be estimated with high sensitivity by analyzing the signals on different spatial modes (e.g., Hermitian Gaussian modes) instead of direct intensity measurements, which has been experimentally demonstrated \cite{39,40,41}. This method has also been generalized to the two- and three- dimensional cases \cite{Pushkina2021, Wang2021}. Recently, Ugo Zanforlin et al. experimentally demonstrated the super-resolution imaging task for resolving two sources with unequal brightness based on hypothesis testing and quantum metrology techniques \cite{33}.  Another resolution-enhanced measurement scheme based on the method of moments has also been proposed, which enables us to analyze and resolve bright unrelated thermal sources without requiring the full measurement statistics \cite{42,43,44}. Although early research mainly focused on incoherent pointlike sources, discrimination of mutually coherent pointlike sources has also been studied \cite{45,46}.

In this paper, we study the super-resolution imaging for a pair of pointlike entangled sources using the SPADE based on the method of moments, where the pointlike entangled sources can be generated by injecting quantum states with different different photon statistics into the OPA. We show that the sensitivity of separation estimation in our scheme does not vanish even if the two entangled sources are infinitely close which renders the diffraction limit irrelevant to the problem. We also show that the sensitivity using the entangled sources is significantly better than those using the incoherent and coherent sources and the sensitivity can be enhanced by increasing the squeezing parameter. In addition, we also show that smaller phase difference between the entangled sources is in favor for better sensitivity. The results here can find important applications for the quantum super-resolution imaging and quantum metrology.


The structure of the paper is as follows. In Sec. II, we illustrate the schematic setup and basic principle for the imaging of a pair of pointlike entangled sources. After introducing the method of moments  for estimating sources separation in Sec. III, we investigate in detail the estimation sensitivity of our scheme in Sec. IV. We summarize our results in Sec. V.

\section{Model of sources and imaging system}

\begin{figure*}[tbp]
\label{Fig1} \centering \includegraphics[width=12cm]{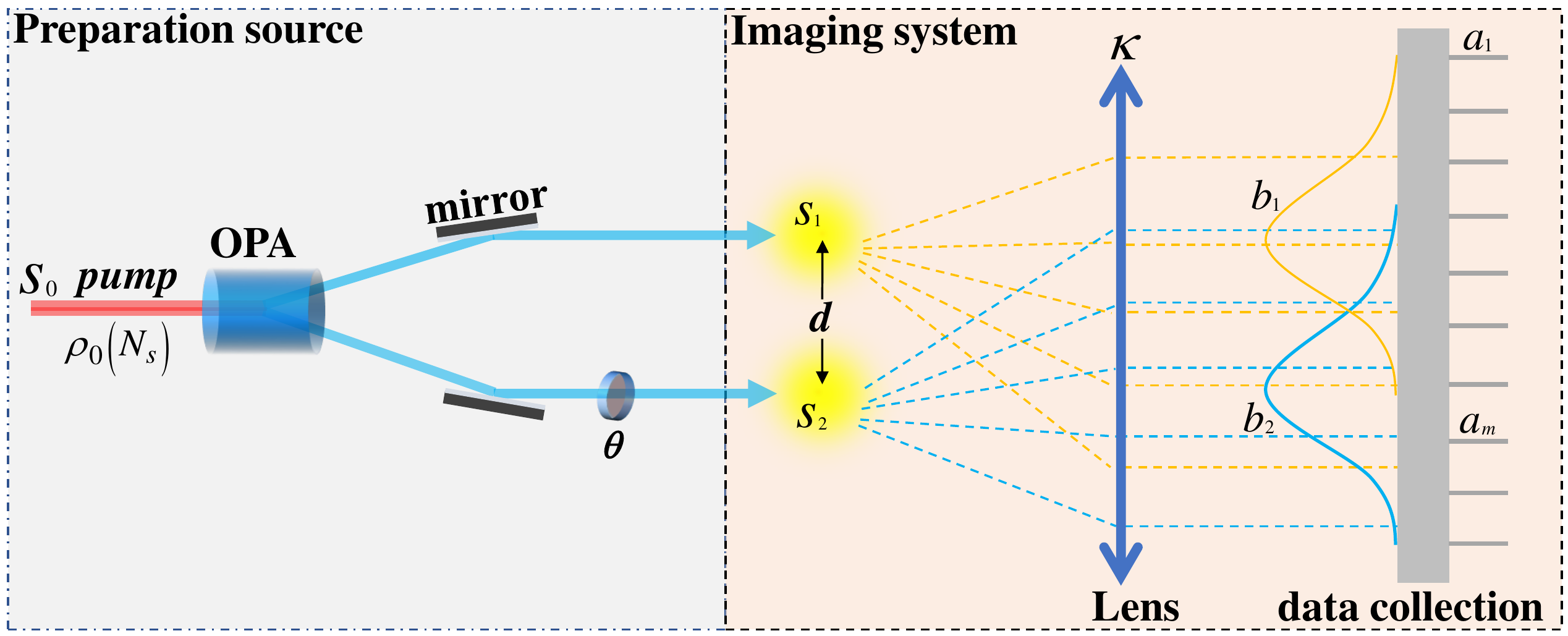}
\caption{{}(Color online) On the left (grey box): conceptual scheme for generating a general two-mode pointlike entangled light sources based on an optical parametric amplification with squeezing parameter $r$ and a phase shifting element $\theta$. On the right (orange box): optical scheme for the estimation of the separation of the sources, where $\kappa $ is a transmissivity of the imaging system, and photon counting is performed in the measurement modes $h_{m}\left( \overrightarrow{r}\right) $ with corresponding field operators $\hat{a}_{m}$.}
\end{figure*}
In this section, we propose theoretically an optical scheme for resolving a pair of pointlike entangled sources (corresponding to the orthogonal modes with field operators $\hat{s}_{1,2}$) located at positions $\mathbf{r}_{1,2}=\left( \pm d/2,0,0\right) $. The light emitted by this sources can be considered the result of a single mode light source that inject into the optical parametric amplification (OPA), and then adds a phase shifting element $\theta $ to one of the output modes (see the left part of Fig.1). The OPA can be described as a two-mode unitary squeezed operator $S_{2}\left(r\right) =\exp \{r\left( \hat{s}_{0}^{\dagger }\hat{v}_{0}^{\dagger }-\hat{s}_{0}\hat{v}_{0}\right) \}$ with a real squeezing parameters $r$. In these situations, the transition of the field operators from the incident field $\hat{s}_{0}$ to the modes of sources $\hat{s}_{1,2}$ can be given by \cite{47}%
\begin{equation}
\left(
\begin{array}{c}
\hat{s}_{1} \\
\hat{s}_{2}^{\dagger}%
\end{array}%
\right) =\left(
\begin{array}{cc}
1 & 0 \\
0 & e^{i\theta }%
\end{array}%
\right) \left(
\begin{array}{cc}
\cosh r & -\sinh r \\
-\sinh r & \cosh r%
\end{array}%
\right) \left(
\begin{array}{c}
\hat{s}_{0} \\
\hat{v}_{0}^{\dagger}%
\end{array}%
\right),   \label{1}
\end{equation}%
where $\hat{v}_{0}^{\dagger}$ is the creation field operator of the vacuum mode. For any initial incident field (a single mode light $\rho _{\hat{s}_{0}}$), corresponding to field operator $\hat{s}_{0}$, the first-order coherency matrix of the modes $\hat{s}_{1,2}$ can be calculated as
\begin{equation}
\left\langle\hat{s}_{i}^{\dagger }\hat{s}_{j}\right\rangle=\left(
\begin{array}{cc}
N_{s_1} & C \\
C^* &
N_{s_2}%
\end{array}%
\right), \label{2}
\end{equation}%
with
\begin{gather}
N_{s_{1}}=N_{s}\cosh ^{2}r+\sinh ^{2}r, \tag*{2(a)} \\
N_{s_{2}}=N_{s}\sinh ^{2}r+\sinh ^{2}r, \tag*{2(b)} \\
C=-\frac{1}{2}e^{i\phi }\text{Tr}[\rho _{s_{0}}s_{0}^{2}]\sinh 2r, \tag*{2(c)}
\end{gather}
where $N_{s}=\text{Tr}[\rho _{0}\hat{s}_{0}^{\dagger }\hat{s}_{0}]$ is the average intensity of the incident field. In particular,the case of $r\rightarrow \infty $ corresponds to equally bright sources and $r\rightarrow 0$ corresponds to all light in mode $\hat{s}_{1}$. Although theoretically we can achieve an arbitrary large squeezing parameter, it is an impossible task experimentally. Therefore, we only focus on finite squeezing parameter $r\in \left[ 0,1\right] $ of the OPA in the following discussion, as far as the current technology is concerned, which means that it is easy to generate entangled light sources experimentally for finite squeezed intensity \cite{48,49}.

In the imaging system, the separation parameters $d$ of the sources $\hat{s}_{1,2}$, are estimated from measurements of the diffracted light. Now, let us consider that a pair of pointlike entangled sources emit light, which passing through a diffraction-limited imaging system with finite aperture that has a transmissivity $\kappa $ and a point spread function (PSF) $u_{0}\left(\mathbf{r}\right) $ (see the right part of Fig. 1). In general, for the case of the paraxial approximation, the parameter $\kappa $ does not depend on the positions of the light source $\hat{s}_{1,2}$. The evolution of the field operators through a diffraction-limited imaging system can actually be described as the following transformations \cite{28}
\begin{equation}
\hat{s}_{1,2}\rightarrow \sqrt{\kappa }\hat{b}_{1,2}+\sqrt{1-\kappa}\hat{v}_{1,2}, \label{3}
\end{equation}
where $\hat{b}_{1,2}=\int d^3\mathbf{r}u(\mathbf{r}-\mathbf{r}_{1,2})\hat{b}_{\mathbf{r}}$ are the image operators and $\hat{v}_{1,2}$ are the field operators of auxiliary environmental modes of diffraction-limited imaging system, which are assumed to be in the vacuum state. Due to the diffraction limit, the image profile functions $u(\mathbf{r}-\mathbf{r}_{1})$ and $u(\mathbf{r}-\mathbf{r}_{2})$ are usually overlapped when the distance between the two point sources is small. Hence, it should be noted that the image operators $\hat{b}_{1}$ and $\hat{b}_{2}$ do not satisfy the usual canonical commutation relations.

In order to obtain the ultimate sensitivity of separation parameters $d$ estimation, we can apply the moment-based estimation technique and here we use the SPADE method to measure the light in the image plane \cite{27,42}. The SPADE can be described as measurements over some field modes $h_{m}\left( \mathbf{r}\right) $ with corresponding operator $\hat{a}_{m}$, where $h_{m}\left( \mathbf{r}\right) $ are general nonlocalized modes (such as the Hermite-Gauss modes \cite{42} which can saturate the QFI for the estimation of the separation between equally bright thermal sources). The input-output relation of the field operators of the measurement modes is given by \cite{45}
\begin{equation}
\hat{a}_{m}=A_{m}\hat{s}_{0}+\hat{\widetilde{v}}_{m},  \label{4}
\end{equation}%
where $A_{m}$ are complex coefficients and $\hat{\widetilde{v}}_{m}$ are nonnormalized nonorthogonal combinations of the field operators of vacuum modes that are orthogonal to the mode $\hat{s}_{0}$. By using the SPADE method, the separation parameters $d$ can in principle be estimated from the measured numbers of photon $N_{m}=\left\langle \hat{a}_{m}^{\dagger }\hat{a}_{m}\right\rangle $ in the image plane. The average number of detected photons in the $m$th measurement mode reads%
\begin{equation}
N_{m}=\left\vert A_{m}\right\vert ^{2}N_{s},  \label{5}
\end{equation}%
and for our scheme as shown in Fig. 1, the complex coefficients $A_{m}$ is given by
\begin{eqnarray}
A_{m} &=&\sqrt{\kappa }\int d^{3}\mathbf{r}h_{m}^{\ast }\left(\mathbf{r}\right) [u_{0}\left( \mathbf{r}-\mathbf{r}_{1}\right) \cosh r -u_{0}\left( \mathbf{r}-\mathbf{r}_{2}\right) e^{i\theta }\sinh r].  \label{6}
\end{eqnarray}%
From Eqs. (\ref{5}) and (\ref{6}), we can see that the average photon number in each measurement modes contain the information of source seperation $d$, the phase difference $\theta $ of sourec modes $\hat{s}_{1,2}$, the transmissivity $\kappa $, the PSF $u_{0}\left(\mathbf{r}\right) $, and the shapes of the Hermite-Gauss modes  $h_{m}\left( \mathbf{r}\right) $.

\section{the method of moment and spatial-mode demultiplexing}

In this section, we derive the measurement sensitivity using the method of moments \cite{50,51}. For a given observable $\hat{O}$,  an estimator $\hat{d}$ of the separation parameter $d$ can be obtained from the sample mean $\overline{o} _{\tau }=\sum_{i=1}^{\tau }o_{i}/\tau $ of $\tau $ independent measurements of $\hat{O}$.
After performing enough measurements, i.e. $\tau \gg1$, according to the central limit theorem, $\overline{o}_{\tau }$ presents the normally distribution with the mean value $\left\langle \hat{O}\right\rangle $ and the variance $\left(\Delta \hat{O}\right) ^{2}=\left\langle \hat{O}^{2}\right\rangle -\left\langle \hat{O}\right\rangle ^{2}$. The estimation error of the separation parameter $d$ can be calculated by
\begin{equation}
\left( \Delta d\right) ^{2}=\frac{\left( \Delta \hat{O}\right) ^{2}}{\tau\left( \partial \left\langle \hat{O}\right\rangle /\partial d\right) ^{2}},  \label{7}
\end{equation}%
which also determines the sensitivity of estimating the separation parameter $d$ in the method of moments. According to the Cramer-Rao lower bound
\begin{equation}
\left( \Delta d\right) ^{2}\geqslant \frac{1}{\tau F(d,\hat{O})}, \label{8}
\end{equation}
where $F(d,\hat{O})$ is\ classical Fisher information. Finally, we can obtain the ultimate sensitivity of separation parameter $d$ by calculating the QFI $F_{Q}\left( d\right) $, i.e., maximizing the classical Fisher information over any positive-operator value measure (POVM) $F_{Q}\left( d\right) =\max_{\hat{O}}F(d,\hat{O})$. Assuming that a set of parameters $\left\{ \lambda_{i}\right\} $ is measured, the measurement sensitivity matrix is then given by
\begin{equation}
S_{ij}=\sum_{m,n}\Lambda _{mn}^{-1}\frac{\partial \left\langle \hat{O}\right\rangle _{m}}{\partial \lambda _{i}}\frac{\partial \left\langle\hat{O}\right\rangle _{n}}{\partial \lambda _{j}}, \label{9}
\end{equation}%
with $\Lambda _{mn}=\left\langle \hat{O}_{m}\hat{O}_{n}\right\rangle-\left\langle \hat{O}\right\rangle _{m}\left\langle \hat{O}\right\rangle _{n}$ being the covariance matrix of the observables. It should be noted that this sensitivity matrix is obtained by optimizing the linear combination of the average values $\left\langle \hat{O}\right\rangle _{m}$ of observable measurements $\{\hat{O}_{m}\}$. The covariance of the estimator $\left\{ \hat{\lambda }_{i}\right\} $ is given by the inverse of the sensitivity matrix, i.e.,%
\begin{equation}
cov\left( \hat{\lambda }_{i},\hat{\lambda }_{j}\right) =\frac{1}{\tau }S_{ij}^{-1}.  \label{10}
\end{equation}

In practice, we cannot measure all observables experimentally. Fortunately, the method of moments allows us to avoid estimating parameters from the full photon counting statistics. We take photon number operators $\hat{N}_{m}=\hat{a}_{m}^{\dagger }\hat{a}_{m}$ in the $m$th measurement mode as observables. By using the Sherman-Morrison formula \cite{52,53} and the method of moments, the elements of the inverse covariance matrix in Eq. (\ref{9}) can be then calculated as
\begin{equation}
\Lambda _{mn}^{-1}=\delta _{mn}N_{m}^{-1}-\frac{g^{\left(2\right) }-1}{1+\left( g^{\left( 2\right) }-1\right) N_{D}},  \label{11}
\end{equation}%
where $g^{\left( 2\right) }=\left( \Delta N_{s}^{2}-N_{s}\right) /N_{s}^{2}$ is the degree of second-order coherence of the initial incident field and $N_{D}=\sum_{m}N_{m}$ is the total average photon number in the image plane. In our scheme, we assume that all parameters except the separation parameter $d$ of the pointlike entangled sources are known. In other words, we only need to estimate the parameter $d$, which corresponds to single-parameter estimation. Then, by substituting Eqs. (\ref{5}) and (\ref{11}) into Eq. (\ref{9}), we can obtain the sensitivity, i.e.,
\begin{equation}
\Re_{d}=N_{D}\sum_{m}\frac{1}{\mathcal N_{m}}\left( \frac{\partial\mathcal N_{m}}{\partial d}\right) ^{2}+\frac{1}{\Delta N_{D}^{2}}\left( \frac{\partial N_{D}}{\partial d}\right) ^{2},  \label{12}
\end{equation}
where $\mathcal N_{m}=N_{m}/N_{D}$ is the ratio of the photon number of the $m$th Hermite-Gaussian mode to the total photon number of the image plane, and $\Delta N_{D}^{2}$ is the variance of the total average photon number in the image plane, which can be calcaulate as%
\begin{equation}
\Delta N_{D}^{2}=\sum_{mn}\Lambda _{mn}=N_{D}\left[ 1+\left( g^{\left(2\right) }-1\right) N_{D}\right].   \label{13}
\end{equation}
According to Eq. (\ref{8}), the variance of the separation parameter $d$ can be derived as
\begin{equation}
\left( \Delta d\right) ^{2}=\frac{1}{\tau }\frac{1}{N_{D}\Re _\mathcal N +\Re _{D}},  \label{14}
\end{equation}%
where

\begin{gather}
\Re _\mathcal N=\sum_{m}\frac{1}{\mathcal N_{m}}\left( \frac{\partial\mathcal N_{m}}{\partial d}\right) ^{2},  \tag*{14(a)} \\
\Re _{D}=\frac{1}{\Delta N_{D}^{2}}\left( \frac{\partial N_{D}}{\partial d}\right) ^{2}. \tag*{14(b)}
\end{gather}%
From  Eq. (\ref{14}), we can find that the ultimate sensitivity of the separation parameter only depends on the relative photon number $\mathcal N_{m}$ and the total photon number in image plane $N_{D}$. $\Re _{\mathcal N}$ can be viewed as the sensitivity of the relative intensity measurements (RIM), which only depends on the relative photon number $\mathcal N_{m}$, while  $\Re _{D}$ can be viewed as the sensitivity of the total photon number detection (TPD).

Next, we analyze the sensitivity of separation parameter estimation in details for different initial incident fields. For this purpose, we consider that the spatial field distribution, in the diffraction-limited imaging system, is a Gaussian PSF \cite{27,54}, i.e.,%
\begin{equation}
u_{0}\left( \mathbf{r}\right) =\sqrt{\frac{2}{\pi \omega ^{2}}}\exp\left( -\frac{\left\vert \mathbf{r}\right\vert ^{2}}{\omega^{2}}\right),   \label{15}
\end{equation}%
where $\omega $ is the width of the PSF. For this PSF, the quantum Cramer-Rao bound can be approached by demultiplexing Hermite-Gauss (HG) modes \cite{45}. The measurement HG modes can be defined as
\begin{equation}
h_{m}\left( x,y\right) =\frac{1}{\sqrt{2^{m}m!}}H_{m}\left( \frac{\sqrt{2}x}{\omega }\right) u_{0}\left( \sqrt{x^{2}+y^{2}}\right),   \label{16}
\end{equation}%
where $H_{m}\left( \cdot \right) $ is the Hermite polynomial. For ideal measurement in the HG mode basis, combining Eqs. (\ref{6}), (\ref{15}) and (\ref{16}), the coefficients $A_{m}$ can be calculated as
\begin{equation}
A_{m}=\sqrt{\kappa }\left[ \left( -1\right) ^{m}\cosh r-e^{i\theta }\sinh r\right] G_{m}\left( \frac{d}{2\omega }\right),   \label{17}
\end{equation}%
where $G_{m}(\gamma )=\gamma ^{m}e^{-\gamma ^{2}/2}/\sqrt{m!}$ with $\gamma=d/2\omega $. Inserting Eq. (\ref{17}) into Eq. (\ref{5}), one can achieve the mean photon number in the measurement modes%
\begin{equation}
N_{m}=N_{s}\kappa \left[ \cosh 2r-\left( -1\right) ^{m}\eta\right] G_{m}^{2}\left( \frac{d}{2\omega }\right),   \label{18}
\end{equation}%
with $\eta=\cos \theta \sinh2r$, and the total mean photon number can be calculated as
\begin{equation}
N_{D}=N_{s}\kappa \left( 1+\wp \eta\right),   \label{19}
\end{equation}%
where $\wp $ is the overlap function between the images of the two point sources, which is given by\cite{42}%
\begin{equation}
\wp =\int u_{0}\left( \mathbf{r}-\mathbf{r}_{1}\right) u_{0}\left( \mathbf{r}-\mathbf{r}_{2}\right) d\mathbf{r} =\exp \left( -\frac{d^{2}}{2\omega ^{2}}\right).  \label{20}
\end{equation}
where we have used the PSF shown in Eq. (\ref{15}).

\section{Sensitivity of the separation parameters $d$ estimation}%

\begin{figure}[tbp]
\label{Fig2} \centering \includegraphics[width=0.32\columnwidth]{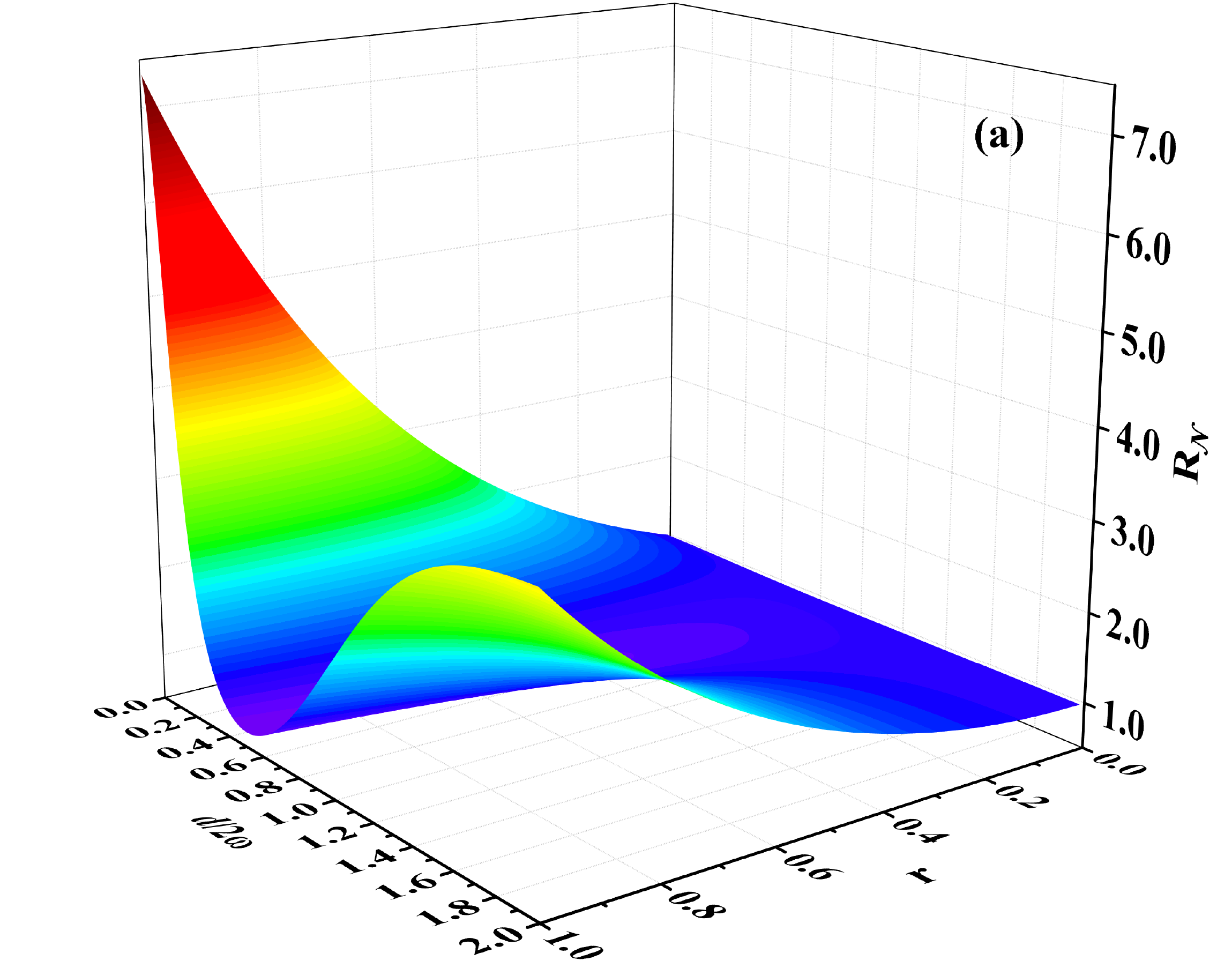} \includegraphics[width=0.32\columnwidth]{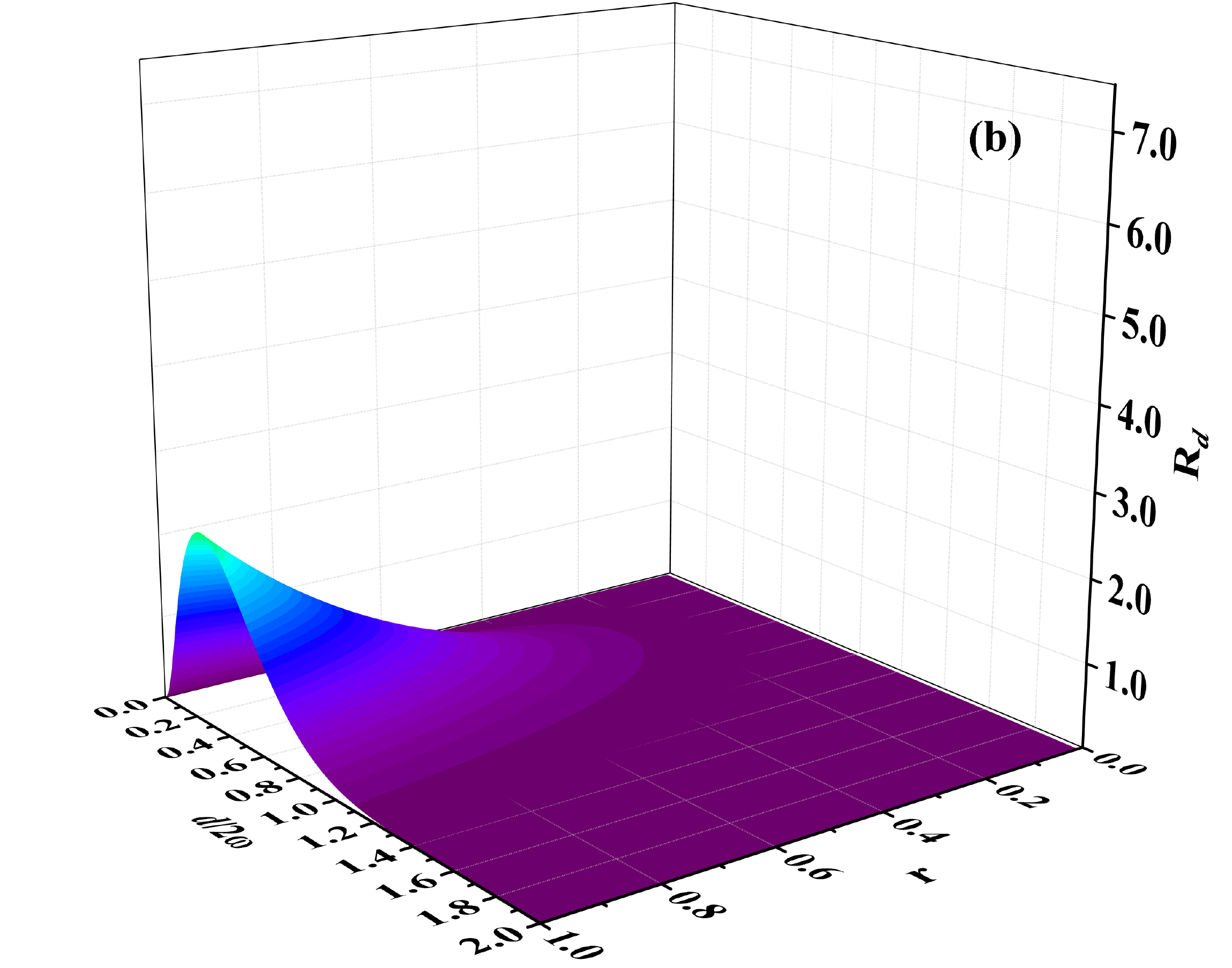}\includegraphics[width=0.32\columnwidth]{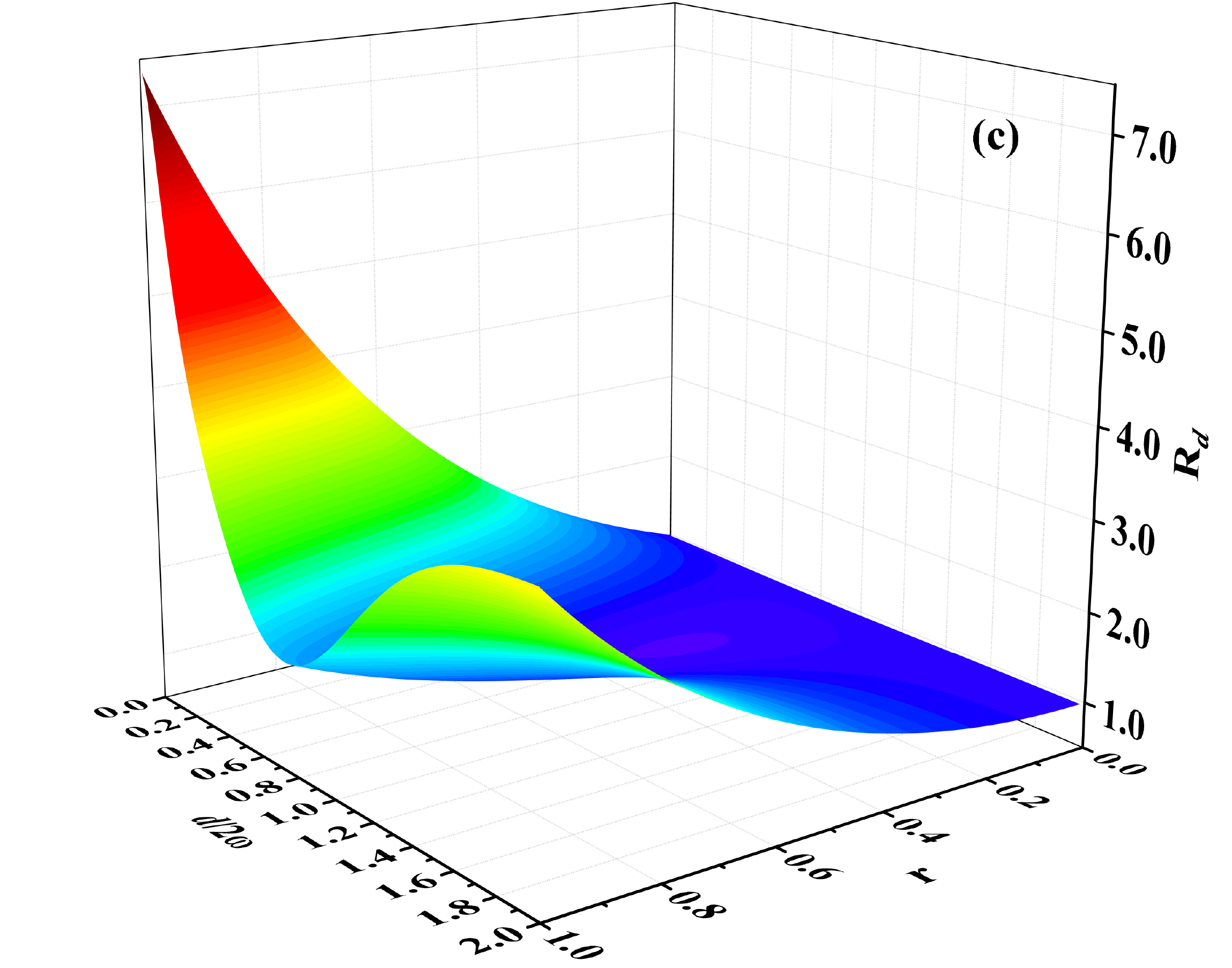}
\caption{{}(Color online) The normalized sensitivity as a function of $r$ and $d/2\omega$ for the symmetrical entangle sources ($\theta=0$). (a) The normalized RIM sensitivity, (b) the normalized TPD sensitivity and (c) the normalized separation estimation sensitivity.}
\end{figure}
In this section, we theoretically and numerically calculate the sensitivity of our scheme under different circumstances. We have demonstrated above that the ultimate sensitivity of parameter $d$ estimation is determined by the RIM and TPD. Here, let's first consider the case when the full HG bases are measured (i.e., $m\rightarrow \infty $). From Eq. 14(a), the RIM sensitivity can be calculated as
\begin{equation}
\Re _{\mathcal N}=\frac{\kappa N_{s}}{\omega ^{2}N_{D}}\left( \cosh 2r+\wp \eta -\frac{d^{2}\wp \eta \cosh 2r}{\omega ^{2}\left( \cosh 2r-\wp \eta \right) }\right). \label{21}
\end{equation}%
For simplicity, we here define $R_{\mathcal N}=\omega^{2}N_{D}\Re _{\mathcal N}/N_{s}\kappa $ as the normalized RIM sensitivity which is given by %
\begin{equation}
R_{\mathcal N}=\cosh 2r+\wp \eta -\frac{d^{2}\wp \eta \sinh 2r}{\omega ^{2}\left( \cosh2r-\wp \eta \right) }.  \label{22}
\end{equation}%
From the above equation, we can see that the normalized RIM sensitivity depends on the source seperation $d$, the phase difference $\theta $, the squeezed parameters $r$ of the OPA but not on the property of the initial incident field $\rho _{0}$.%

\begin{figure}
\label{Fig3} \centering \includegraphics[width=0.4\columnwidth]{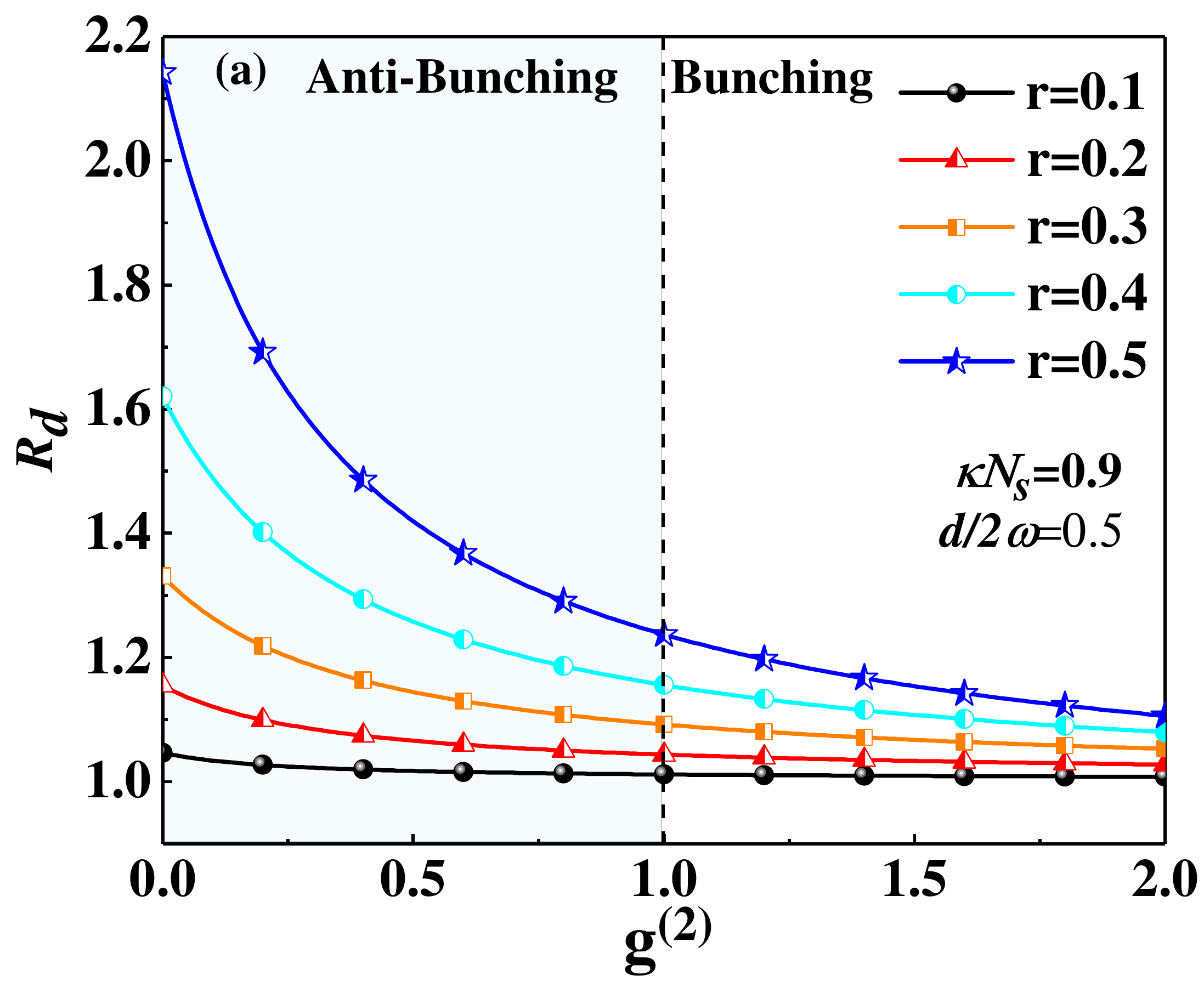}\includegraphics[width=0.4\columnwidth]{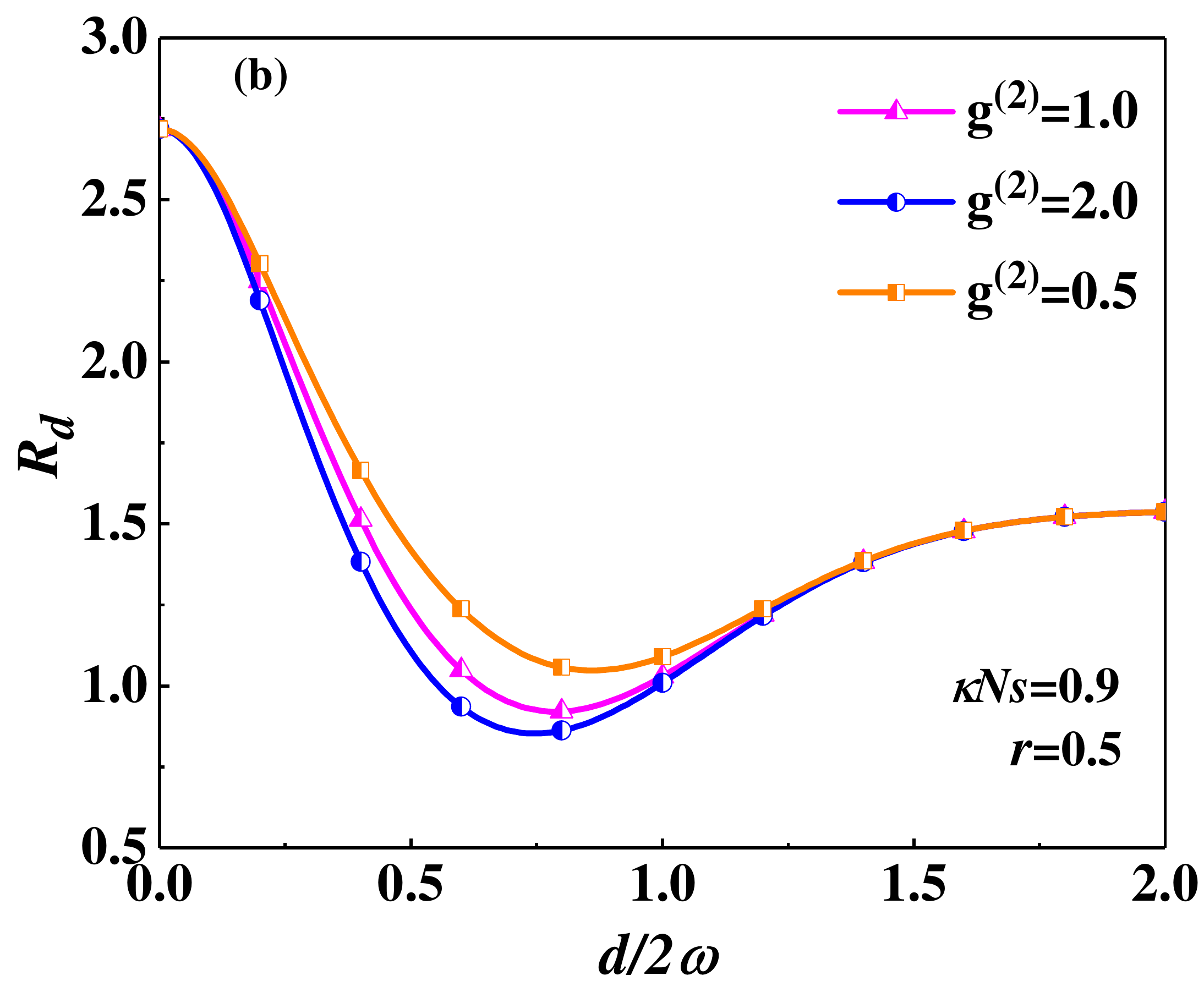}
\caption{{}(Color online)  The normalized separation estimation sensitivity as a function of the second-order coherence degree $g^{\left( 2\right) }$ for a given combined value  $N_{s}\kappa=0.9 $ of the transmittance of the diffraction limit system and the average photon number of the initial incident field. (a) The separation parameter $d/2\omega=0.5$ with the different of squeezed parameter, (b) the squeezed parameter $r=0.5$ with the different of the initial incident field. The blue region represents that the  non-classical light as the initial incident field, while the white region represents classical light.}
\end{figure}
On the other hand, for the TPD sensitivity, it is seen from Eq. 14(b) that the TPD sensitivity does not depend on average  photon number of individual HG modes $N_{m}$ while only depends on the total average photon number detection $N_{D}$. Substituting Eqs. (\ref{13}) and (\ref{19}) into 14(b), we can obtain
\begin{equation}
\Re _{D}=\frac{N_{s}^{2}\kappa ^{2}d^{2}\wp ^{2}\eta ^{2}}{\omega ^{4}N_{D}\left[ 1+N_{s}\kappa \left( g^{\left( 2\right) }-1\right) \left( \cosh2r-\wp \eta \right) \right] }.  \label{23}
\end{equation}%
It is not difficult to find that the TPD sensitivity depends on the quantum statistics distribution of the incident light field. The incident light field with antibunching statistics $\left( g^{\left( 2\right) }<1\right) $ can provide a better sensitivity than those with bunching statistics $\left( g^{\left( 2\right)}>1\right) $. This is because the incident field with antibunching statistics has a lower photon number variance. Similar to the RIM sensitivity, here we also define $R_{D}=\omega ^{2}\Re _{D}/N_{s}\kappa $ as the normalized TPD sensitivity, which is given by%
\begin{equation}
R_{D}\text{=}\frac{\wp ^{2}\eta ^{2}d^{2}(\cosh 2r-\eta)^{-1}}{2\omega ^{2} \left[ 1+\left( g^{\left( 2\right) }-1\right) N_{s}\kappa \left(\cosh 2r-\wp \eta \right) \right] }.  \label{24}
\end{equation}

Finally, we discuss overall sensitivity of the separation $d$ estimation, which is given by Eq. (\ref{14}) and its normalized sensitivity can be obtained by combining the RIM sensitivity Eq. (\ref{23}) with the TPD sensitivity Eq. (\ref{24}), i.e.,
\begin{equation}
R_{d}=R_{\mathcal N}+R_{D}  \label{25}
\end{equation}
We assume that the incident field $\rho_{0}$ is in a coherent state whose photon statistics is Poisson distributed and $g^{\left( 2\right) }=1$. In this case, we have
\begin{equation}
R_{D}=\frac{\wp ^{2}\eta ^{2}d^{2}}{2\omega ^{2}\left( \cosh 2r-\eta \right)}.  \label{26}
\end{equation}
From Eqs. (\ref{25}) and (\ref{26}), we can find that, after propagation through the diffraction-limited imaging system, the normalized separation estimation sensitivity does not depend on the source intensity of the coherent state.%

\begin{figure}
\label{Fig4} \centering\includegraphics[width=0.5\columnwidth]{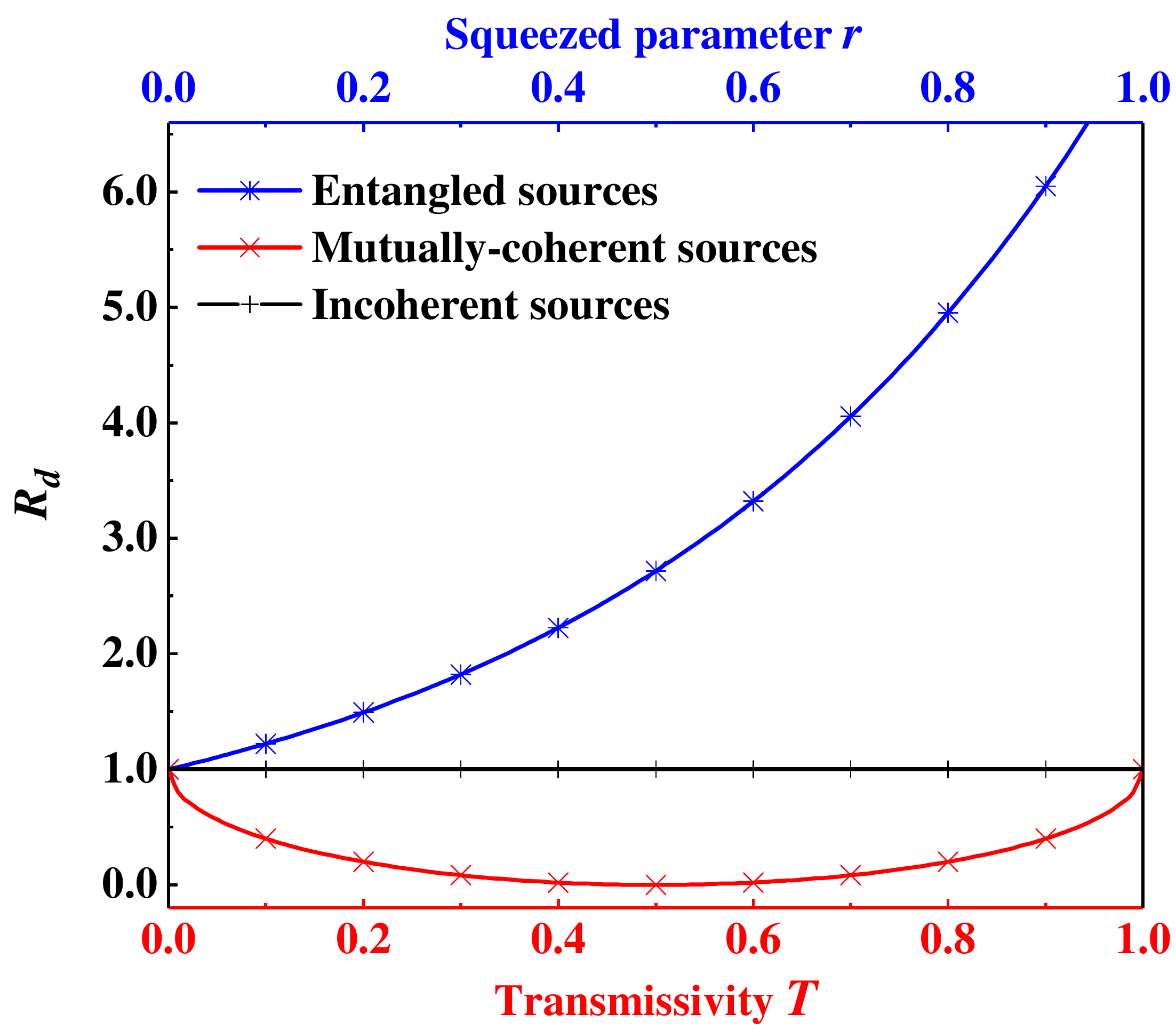}
\caption{{}(Color online) The asymptotic behavior ($d\to0$) of the normalized separation estimation sensitivity for the entangled sources (Blue with stars line), the incoherent sources (Black with bars line), and the mutually-coherent sources (Red with cross line).}
\end{figure}

In order to clearly see the effect of different parameters $d$ and $r$ on the amount of the normalized sensitivity, we plot the normalized sensitivity $R_{\mathcal N}$ as a function of $r$ and $d$ when $\theta=0$ (Fig. 2). Form Fig. (2a), we can see that the improvement of the normalized RIM sensitivity for a fixed $d$ can be found by increasing the squeezed parameters $r$ of the OPA especially when $d$ approaches zero. Surprisingly, the optimal value of the normalized RIM sensitivity appear in the range of $d\ll2\omega $ which is completely different from that using the direct imaging  where the sensitivity approach zero for very small separation (Fig. 2(b)). This indicates that the pointlike entangled symmetric sources can still be distinguished by using the SPADE method even when the separation $d$ is much less than the diffraction limit. Meanwhile, we also find that the ultimate separation $d$ estimation sensitivity of symmetric sources is mainly dependent on the RIM sensitivity (Fig. (2c)). This is due to the fact that the TPD sensitivity of the coherent state is far less than the RIM sensitivity for a certain parameter $(d,r)$ threshold, i.e., $R_{\mathcal N}\gg R_{D}$. 


On the other hand, we also study the separation estimation sensitivity for different second-order correlation $g^{\left( 2\right) }$ of the initial incident light field. From Eq. (24), we can see that the quantum statistics of the incident field can affect the separation estimation sensitivity. From Fig. 3(a), we can see that the separation estimation is enhanced when $g^{\left( 2\right) }$ decreases for all the chosen squeezing parameters. This indicates that anti-bunching of the incident field is beneficial for the separation estimation when $d/2\omega=0.5$. In Fig. 3(b), we compare the separation estimation sensitivity as a function of $d$ for the coherent state ($g^{\left( 2\right) }=1$), the thermal state ($g^{\left( 2\right) }=2$) and the antibunching state ($g^{\left( 2\right)}=0.5$) as the initial incident sources. We can find that the light sources with different $g^{\left( 2\right)}$ can give different estimation sensitivities when $d\approx \omega$ and antibunching light source can improve the sensitivity in this region. However, the separation estimation sensitivity of the pointlike entangled sources generated by the non-classical source is not significantly improved than that of the classical sources when the separation $d$ is much less than the diffraction limit (i.e., $d\rightarrow 0$). Therefore, when the source separation is very small, the pointlike entangled sources generated by the OPA can give similar sensitivity for any initial incident sources in our scheme. %
\begin{figure}
\label{Fig5} \centering\includegraphics[width=0.35\columnwidth]{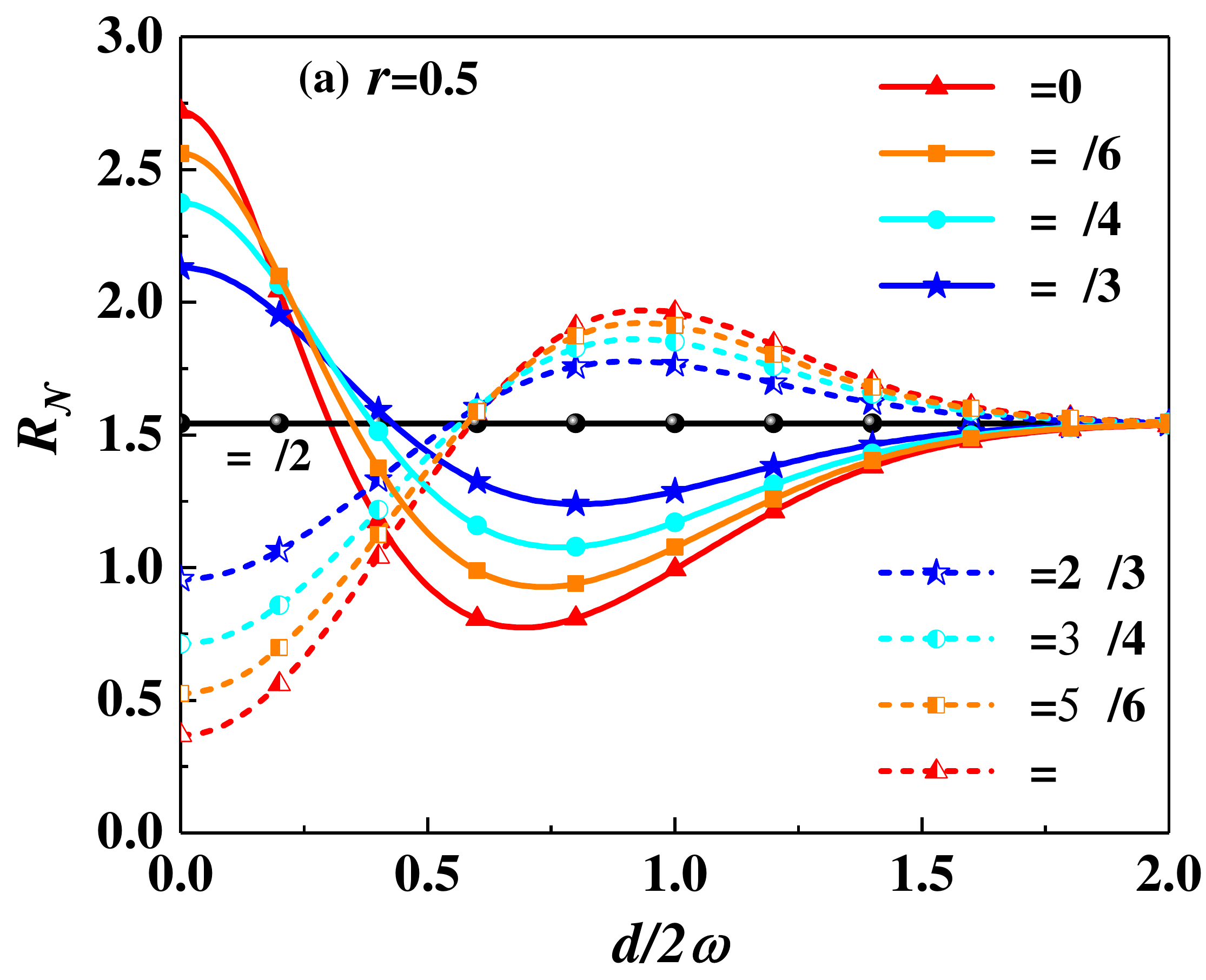}\includegraphics[width=0.35\columnwidth]{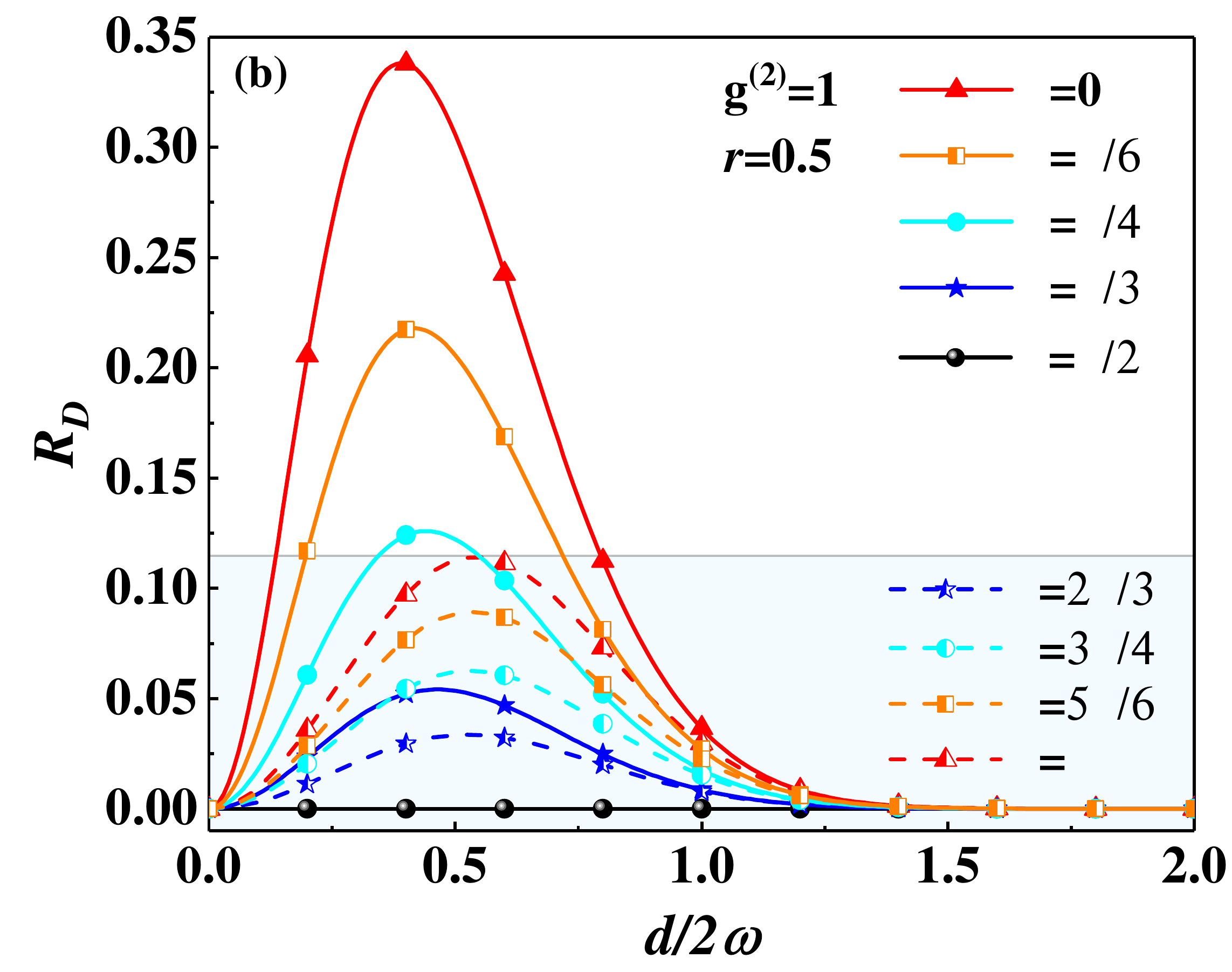}\includegraphics[width=0.35\columnwidth]{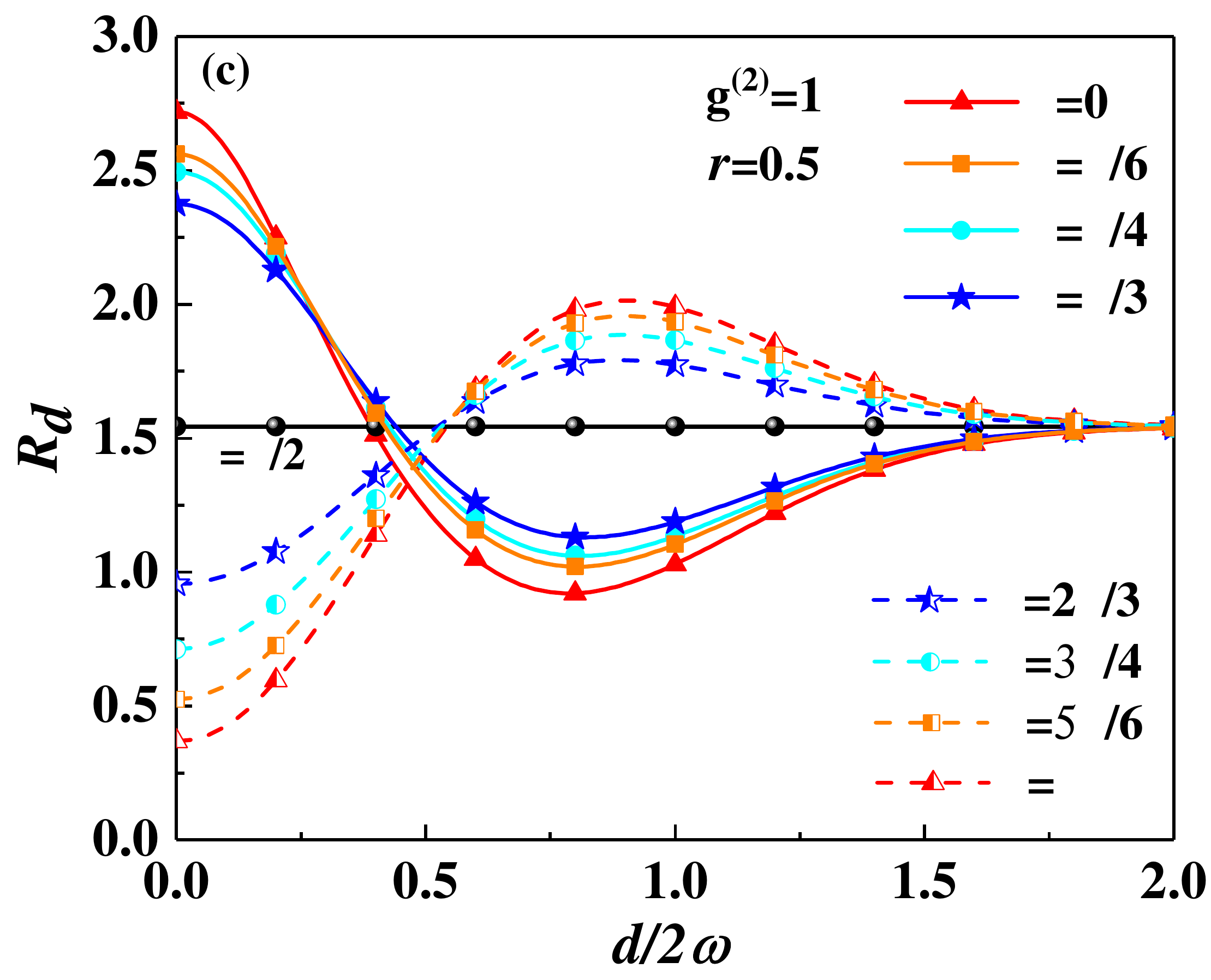}
\caption{{}(Color online) The normalized sensitivity as a function of $r$ and $d/2\omega$ for the symmetrical entangle sources ($\theta=\pi$). (a) The normalized RIM sensitivity, (b) the normalized TPD sensitivity and (c) the normalized separation estimation sensitivity.}
\end{figure}

In Fig. 4, we compare the normalized separation estimation sensitivities for the incoherent sources\cite{42,43}, mutually-coherent sources \cite{45}, and entangled sources when $d\to0$. From the figure, we can see that the normalized separation estimation sensitivity $R^{inc}_d=1$ for the incoherent sources and $R^{muc}_d=1-2\sqrt{T(1-T)}$ for mutually coherent sources generated by the optical beam splitter, where $T=\cos^2\phi$ is the transmissivity of the optical beam splitter used in Ref. \cite{45}. It is seen that the estimation sensitivity of the coherent sources is less than that of the incoherent sources, and interestingly the normalized sensitivity disappears for the equally bright mutually coherent sources ($T=0.5$).  In contrast, the estimation sensitivity of the entangled sources is always greater than 1 and increases as $r$ increases (blue line with star symbols in Fig. 4). This can be easily seen from the asymptotic behavior of the estimation sensitivity when $d\to0$. 
When $d\to0$ and $\theta=0$, $R_D\to0$ which can be seen from Eq. (26) and $R_d\approx R_{N}\to\cosh2r+\sinh2r$ which is always larger than 1 and clearly increases with $r$.  These results indicate that the estimation sensitivity can be enhanced when the entangled light sources are used instead of coherent and incoheren classical sources.


Finally, in Fig. 5, we discuss the impact of the phase differences $\theta \in \lbrack 0,\pi ]$ on the normalized sensitivity when given squeezed parameters $r=0.5$ and $g^{(2)}=1$. Our numerical results show that increasing the phase difference $\theta$ of the pointlike entangled sources can reduce the RIM sensitivity of our scheme when $d\ll 2\omega $, but it can increase the sensitivity when $d\simeq 2\omega $ [see Fig. 5(a)]. For TPD  sensitivity,  when $0\leq\theta\leq\pi/2$, $R_D$ decreases as $\theta$ increases. In contrast, when $\pi/2\leq\theta\leq\pi$, $R_D$ increases as $\theta$ increases. When $\theta=\pi/2$, $R_D=0$. The optimal normalized TPD sensitivity of resolving symmetric sources ($\theta=0$) is about $1.75$ times higher than that of antisymmetric sources ($\theta=\pi$). From Fig. 5(c), when $\theta =\pi /2$, $R_{d}=R_{\mathcal N}=\cosh 2r/\omega ^{2}$ which does not depend on $d$. For the small separation (i.e., $d\ll2\omega $), $R_d$ increases as $\theta$ decreases and when $\theta=0$ (symmetrical entangled sources) we have the largest measurement sensitivity. In the contrast, when $d\approx 2\omega$, $R_d$ is the largest when $\theta=\pi$. These restuls indicate that for deep subdiffraction limit separation, the entangled sources with zero phase difference can have the largest estimation sensitivity.

\section{Conclusion}
In summary, we analyze the separation estimation sensitivity of pointlike entangle sources, which is produced by injecting  a single-mode light source with arbitrary quantum statistics distribution into an OPA. By using the method of moments together with the SPADE, we show that the separation estimation sensitivity is completely determined by the photon distribution in each detected modes and the total photon number in the image plane. The results show that the separation estimation error does not diverge even when $d\rightarrow 0$ when the light sources are the pointlike entangled sources generated by the OPA. This indicates that the Rayleigh's curse can be overcome in our scheme. Moreover, the detection sensitivity in the case of symmetric entangled light sources can increase with the squeezing parameter $r$ when $d\rightarrow 0$. In addition,
 we also compare the effects of the quantum statistical distribution of different incident light sources on the realization of super-resolution imaging. We find that for the separation around the diffraction limit, the incident light field with antibunching photon statistics has higher measurement sensitivity, but when the separation is much smaller than the diffraction limit,  the pointlike entangled light source generated by OPA with arbitrary initial incident light field, can achieve similar measurement sensitivity in our scheme. The results here can find important applications in the quantum imaging and metrology with super-resolution and super-sensitivity.


\section{Acknowledgments}

This work was supported by the National Key R$\&$D Program of China (Grant No. 2021YFA1400800), the Key-Area Research and Development Program of Guangdong Province (Grant No.2018B030329001), the Guangdong Special Support Program (Grant No.2019JC05X397), the Guangdong Basic and Applied Basic Research Foundation (Grant No. 2023B1515040023), and the Natural Science Foundations of Guangdong (Grant No.2021A1515010039). Wei Ye is supported by the Scientific Research Startup Foundation (Grant No. EA202204230) at Nanchang Hangkong University.


\end{document}